\documentclass[acmsmall,screen]{acmart}
\AtBeginDocument{%
  }

\usepackage{multirow}
\usepackage{fontawesome}
\usepackage{hyperref}
\usepackage{algorithm}
\usepackage{algpseudocode}
\usepackage{amsmath}
\usepackage[most]{tcolorbox}
\usepackage{xcolor}
\newtcolorbox{AnswerBox}{
  colback=gray!20,
  colframe=black,
  arc=2mm,
  boxrule=1.0pt,
  left=6pt,
  right=6pt,
  top=6pt,
  bottom=6pt,
}

\usepackage{xspace}
\newcommand{\methodname}{\textsc{Agent-CoEvo}\xspace}

\setcopyright{acmlicensed}
\copyrightyear{2025}
\acmYear{2025}
\acmDOI{XXXXXXX.XXXXXXX}
\acmISBN{978-1-4503-XXXX-X/2018/06}

\begin{document}

\title{Beyond Fixed Tests: Repository-Level Issue Resolution as Coevolution of Code and Behavioral Constraints}


\author{Kefan Li}
\author{Yuan Yuan}
\affiliation{%
  \institution{Beihang University}
  \city{Beijing}
  \country{China}
}
\affiliation{%
  \institution{Beijing Tokfinity Technology Co., Ltd.}
  \city{Beijing}
  \country{China}
}

\author{Mengfei Wang}
\author{Shihao Zheng}
\author{Wei Wang}
\author{Ping Yang}
\affiliation{%
  \institution{Beijing Tokfinity Technology Co., Ltd.}
  \city{Beijing}
  \country{China}
}

\author{Mu Li}
\author{Weifeng Lv}
\affiliation{%
  \institution{Beihang University}
  \city{Beijing}
  \country{China}
}

\renewcommand{\shortauthors}{Li et al.}

\begin{abstract}
Software engineers resolving repository-level issues do not treat existing tests as immutable correctness oracles. Instead, they iteratively refine both code and the tests used to characterize intended behavior, as new modifications expose missing assumptions or misinterpreted failure conditions. In contrast, most existing large language model (LLM)-based repair systems adopt a linear pipeline in which tests or other validation signals act mostly as post-hoc filters, treating behavioral constraints as fixed during repair. This formulation reduces repair to optimizing code under static and potentially misaligned constraints, leading to under-constrained search and brittle or overfitted fixes.
We argue that repository-level issue resolution is fundamentally not optimization under fixed tests, but search over evolving behavioral constraints. To operationalize this view, we propose Agent-CoEvo, a coevolutionary multi-agent framework in which candidate code patches and test patches are jointly explored and iteratively refined. Rather than treating tests as immutable oracles, our framework models them as dynamic constraints that both guide and are revised by the repair process. Through mutual evaluation and semantic recombination, code and test candidates progressively narrow the space of behavior consistent with the issue description.
Evaluated on SWE-bench Lite and SWT-bench Lite, Agent-CoEvo consistently outperforms state-of-the-art agent-based and agentless baselines in both repair success and test reproduction quality. Our findings suggest that enabling repair agents to revise behavioral constraints during search is critical for reliable issue resolution, pointing toward a shift from code-only optimization to coevolution of implementation and specification.
\end{abstract}

\maketitle

\section{Introduction}

Large language models (LLMs) have demonstrated strong capabilities across a wide range of software engineering tasks, 
including code generation \cite{jiang2024survey,chen2021evaluating,austin2021program,li2022competition}, program repair \cite{xia2023automated,olausson2023self,shinn2023reflexion}, and test construction \cite{schafer2023empirical,kang2023large, nie2023learning}. These advances have spurred growing interest in automating repository-level issue resolution, 
where the goal is to diagnose a behavioral discrepancy described in an issue report and synthesize a code modification that integrates correctly into a complex software system \cite{freund1997decision,jimenez2023swe,guo2025omnigirl}. 
Compared with function-level tasks \cite{chen2021evaluating,austin2021program}, repository-level issue resolution requires reasoning over multi-file dependencies, project-specific invariants, and dynamic execution behaviors.

In real-world software development, resolving an issue rarely relies on fixed and fully reliable 
correctness oracles \cite{barr2014oracle}. Instead, engineers iteratively refine both code and the tests used to characterize intended behavior. 
As new code modifications are attempted, previously hidden assumptions, incomplete specifications, or misinterpreted failure conditions are often exposed, 
prompting revisions to existing tests or the creation of new ones. In this process, tests serve not as immutable definitions of correctness, but as evolving approximations 
of the intended behavioral constraints.

Most existing automated repair systems, however, are built upon a different modeling assumption. Repository-level repair pipelines typically treat available validation signals as fixed during repair: candidate patches are generated first and subsequently filtered based on whether they satisfy these signals. Representative approaches, including Agentless~\cite{xia2024agentless}, SpecRover~\cite{ruan2024specrover}, Moatless Tools~\cite{orwall2024moatless}, KGCompass~\cite{ma2025thinking}, and DARS~\cite{aggarwal2025dars}, largely follow this paradigm, using tests primarily as static acceptance criteria rather than as variables that evolve during search. Under this formulation, repair effectively reduces to optimizing code under fixed validation conditions, implicitly treating behavioral constraints as stable throughout the repair process.

This assumption introduces a tension with the nature of real-world issue resolution.
From a problem-solving perspective, 
repair is a search over candidate patches that requires constraints to evaluate solution quality. 
In practice, these constraints are largely instantiated by tests, but the tests used in automated pipelines are often generated artifacts derived from partial understanding of the issue and may themselves contain errors or omissions.
Treating such tests as fixed constraints forces the search to operate under unreliable assumptions: many incorrect patches can 
satisfy an incomplete test set (under-constrained search), while correct fixes may be rejected due to flawed or mis-specified tests. 
As a result, repair systems may produce brittle or overfitted solutions that satisfy the given tests without faithfully addressing the underlying behavioral discrepancy.

We therefore argue that repository-level issue resolution should be understood not as optimization under fixed tests, 
but as search over evolving behavioral constraints. In this view, tests are not merely auxiliary artifacts 
for post-hoc validation, but revisable constraints that should be refined together with candidate code patches. 
This reframing shifts the focus from code-only optimization to a coupled search process over both implementation and specification.

To realize this perspective, we propose \textbf{\methodname}, a coevolutionary multi-agent 
framework that jointly explores and refines code patches and test patches. Instead of treating tests as immutable oracles, 
our framework models them as dynamic constraints that both guide and are revised by the repair process. 
Through mutual evaluation and semantic recombination, candidate code and test artifacts progressively narrow the space of behaviors consistent with the issue description.
This design naturally enables the system to produce both code patches and corresponding test patches within the same search process, eliminating the traditional separation between repair and test generation.

We evaluate \methodname\ on SWE-bench Lite \cite{jimenez2023swe} and SWT-bench Lite \cite{mundler2024swt}, two complementary benchmarks derived from the same set of real-world GitHub issues. 
SWE-bench Lite measures the correctness of repository-level code repairs, while SWT-bench Lite evaluates the ability to construct tests that faithfully reproduce the issue behavior. 
This shared-issue setup enables a unified evaluation of both repair quality and the refinement of behavioral constraints within a single framework.
We compare \methodname\ against state-of-the-art agent-based and agentless baselines under comparable model settings. 
Experimental results show that \methodname\ achieves consistent improvements in both repair success and test reproduction quality, 
supporting the effectiveness of modeling code and tests as coupled components of the same search process.

Our contributions are summarized as follows:
\begin{itemize}
\item \textbf{Problem Reframing.}  We identify the fundamental mismatch between real-world issue resolution and the fixed-test assumption, and formalize repository-level repair as search under uncertain and evolving behavioral constraints.
\item \textbf{Scalable Coevolution Framework.} We propose a repository-level coevolution framework that makes search under evolving behavioral constraints feasible in the presence of complex project dependencies.
The framework is operationalized through interacting agents and incorporates cross-population evaluation, semantic recombination, and shared localization mechanisms that enable coordinated refinement of code and tests in large-scale repair settings.
\item \textbf{Empirical Validation.}  We conduct extensive evaluations on SWE-bench Lite and SWT-bench Lite benchmarks. The results show that our approach achieves consistent improvements over state-of-the-art agent-based and agentless baselines under comparable model settings, supporting the effectiveness of modeling code and tests as coupled components of the repair search process.
\end{itemize}

\section{Method}
\subsection{Architecture Overview}

\begin{figure}[h]
\centering
\includegraphics[width=\linewidth]{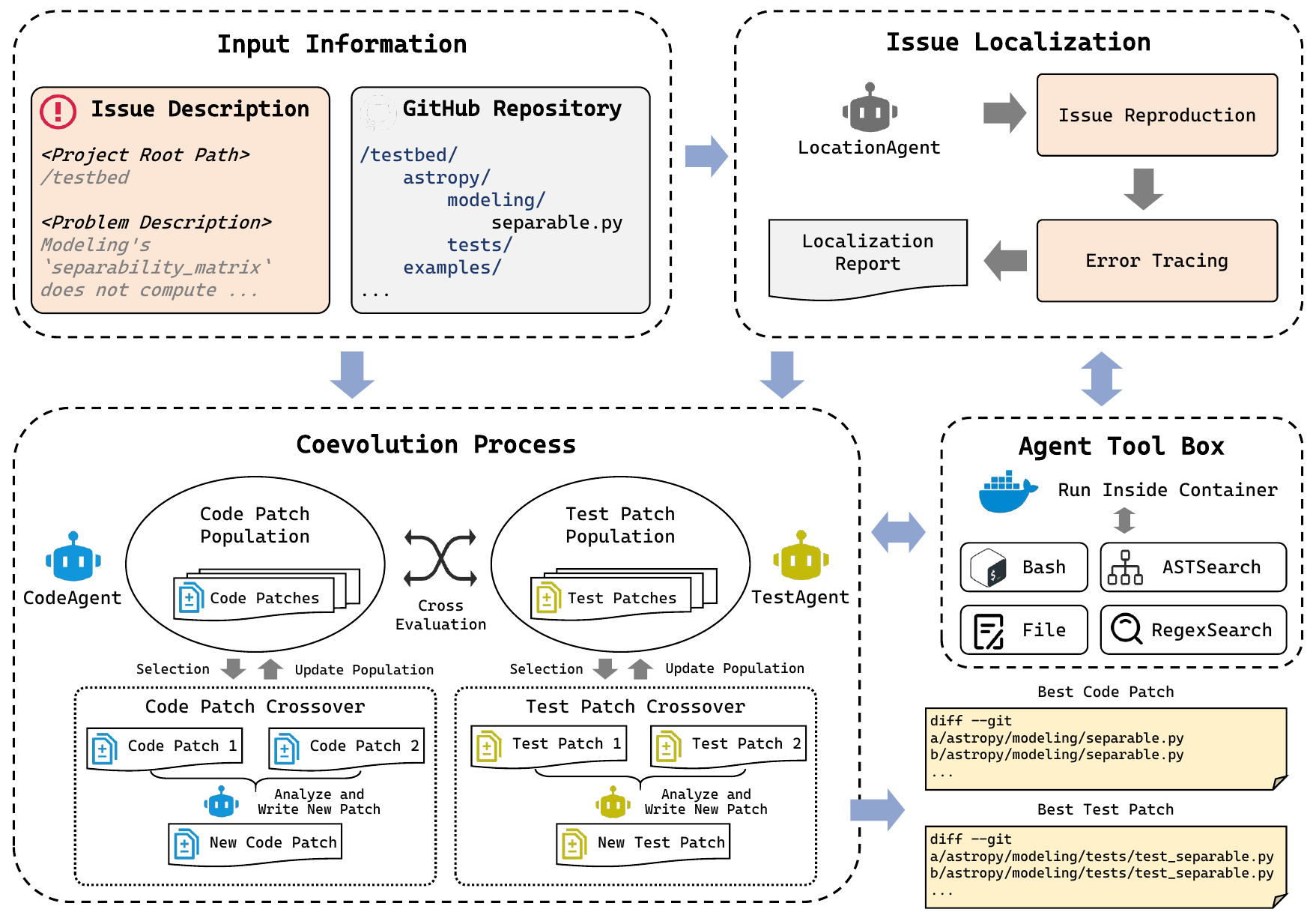}
\caption{
Overview of the \methodname framework. 
}
\label{fig:overview}
\Description{
Overview of
}
\end{figure}

As illustrated in Figure \ref{fig:overview}, we propose \methodname, a coevolutionary repair framework that operationalizes repository-level issue resolution as search under evolving behavioral constraints.
The system takes an issue description $D$ and a GitHub repository $R$ as input, aiming to produce an optimal patch pair $(c^*, t^*)$, where $c^*$ is a code patch and $t^*$ is a test patch.
The framework operates in two distinct stages.
The first stage is \textit{Issue Localization}, which analyzes the unstructured description $D$ to reproduce the reported issue and identify the files and lines in $R$ that are likely responsible for the observed failure.

The second stage is the \textit{Coevolution Process}, which constitutes the core search loop (Algorithm \ref{alg:coevolution-main}).
Two agents, the CodeAgent and the TestAgent, maintain populations of candidate code patches $\mathcal{P}_{\text{code}}$ and test patches $\mathcal{P}_{\text{test}}$, respectively.
Over $K$ iterations, these populations are iteratively refined through mutual evaluation and recombination, forming a coupled search process over implementation hypotheses (code) and behavioral constraint hypotheses (tests).
All agent operations are executed within an isolated Docker environment to ensure repository integrity.

\subsection{Automated Issue Localization}

The LocationAgent serves as the entry point of the framework. 
It processes the natural language issue description $D$ to perform precise fault localization.
Initially, the agent performs action decomposition to synthesize a reproduction script. 
Unlike traditional methods that rely on existing regression tests, the agent constructs new scripts tailored to the failure scenario described in $D$.
Upon execution, the agent generates a dynamic execution trace. 
This trace enables the system to identify the error location at the granularity of specific files and lines.
The output of this stage includes the localized file paths and a root cause analysis, serving as the context for subsequent patch generation.

\subsection{Coevolution of the CodeAgent and TestAgent}
This section details the iterative optimization process.
We model the program repair task as a dual-population evolutionary algorithm.
To optimize computational resources, the localization phase is performed once. 
Subsequently, the framework enters the main evolutionary loop until convergence or the maximum iteration limit $N$ is reached.

\begin{algorithm}[ht!]
\caption{Coevolution Process}
\label{alg:coevolution-main}
\begin{algorithmic}[1]
\Require 
    Population Size $N$, Max Iterations $K$, Issue Description $D$, Repository $R$
\Ensure 
    Best Patch Pair $(c^*, t^*)$

\State \textbf{Phase 1: Initialization}
\State $Ag_{\text{code}} \gets \text{CodeAgent}(R, D, N)$
\State $Ag_{\text{test}} \gets \text{TestAgent}(R, D, N)$

\State $\mathcal{P}_{\text{code}} \gets Ag_{\text{code}}.\text{GenerateInitial}(N)$
\State $\mathcal{P}_{\text{test}} \gets Ag_{\text{test}}.\text{GenerateInitial}(N)$

\State \textbf{Phase 2: Filter Test Patches}
\State $\mathcal{P}_{\text{test}} \gets \{t \in \mathcal{P}_{\text{test}} \mid \texttt{Run}(R, t) = \text{FAIL}\}$ \Comment{Ensure tests fail on bug}

\For{$k \gets 1$ \textbf{to} $K$}
    \State \textbf{Phase 3: Cross Evaluation}
    \State $(F_{\text{code}}, F_{\text{test}}) \gets \Call{CrossEvaluation}{\mathcal{P}_{\text{code}}, \mathcal{P}_{\text{test}}, R}$

    \State \textbf{Phase 4: Elitism \& Selection}
    \State $e_{\text{code}} \gets \arg\max_{c \in \mathcal{P}_{\text{code}}} F_{\text{code}}(c)$
    \State $e_{\text{test}} \gets \arg\max_{t \in \mathcal{P}_{\text{test}}} F_{\text{test}}(t)$
    
    \State $\mathcal{P}^{\text{next}}_{\text{code}} \gets \{e_{\text{code}}\}, \quad \mathcal{P}^{\text{next}}_{\text{test}} \gets \{e_{\text{test}}\}$
    
    \State \textbf{Phase 5: Crossover and Update}
    \While{$|\mathcal{P}^{\text{next}}_{\text{code}}| < N$}
        \State $p_1, p_2 \gets \texttt{TournamentSelect}(\mathcal{P}_{\text{code}}, F_{\text{code}})$
        \State $c_{\text{new}} \gets Ag_{\text{code}}.\text{Crossover}(p_1, p_2)$
        \State $\mathcal{P}^{\text{next}}_{\text{code}} \gets \mathcal{P}^{\text{next}}_{\text{code}} \cup \{c_{\text{new}}\}$
    \EndWhile

    \While{$|\mathcal{P}^{\text{next}}_{\text{test}}| < N$}
        \State $p_1, p_2 \gets \texttt{TournamentSelect}(\mathcal{P}_{\text{test}}, F_{\text{test}})$
        \State $t_{\text{new}} \gets Ag_{\text{test}}.\text{Crossover}(p_1, p_2)$
        \State $\mathcal{P}^{\text{next}}_{\text{test}} \gets \mathcal{P}^{\text{next}}_{\text{test}} \cup \{t_{\text{new}}\}$
    \EndWhile
    
    \State $\mathcal{P}_{\text{code}} \gets \mathcal{P}^{\text{next}}_{\text{code}}, \quad \mathcal{P}_{\text{test}} \gets \mathcal{P}^{\text{next}}_{\text{test}}$
\EndFor

\State \Return Pair with highest fitness in $\mathcal{P}_{\text{code}}, \mathcal{P}_{\text{test}}$
\end{algorithmic}
\end{algorithm}

\subsubsection{Population Initialization}

The initialization phase establishes the starting state for evolution.
The agents independently generate initial populations $\mathcal{P}_{\text{code}}$ and $\mathcal{P}_{\text{test}}$ of size $N$ using LLMs.
For test patches, we retain only those that fail on the buggy repository $R$:
\begin{equation}
\mathcal{P}_{\text{test}} \leftarrow \{t \in \mathcal{P}_{\text{test}} \mid \texttt{Run}(R, t) = \text{FAIL}\}.
\end{equation}
This filtering ensures that generated tests are behaviorally relevant to the reported issue.
Importantly, this step does not assume that tests are complete or fully correct specifications; rather, their adequacy is progressively refined during coevolution.

\subsubsection{Cross-Evaluation Mechanism}

The fitness evaluation relies on the interaction between the code and test populations, as detailed in Algorithm \ref{alg:cross-eval}. 
This process consists of constructing an execution matrix and computing consistency-based fitness.

Conceptually, the execution matrix can be interpreted as an evolving constraint map. Tests impose behavioral constraints over code hypotheses, 
while code behavior provides feedback on the informativeness and adequacy of tests. 
Through iterative updates, this interaction progressively narrows the space of behaviors consistent with both the issue description and observed executions, 
enabling a coupled refinement of implementation and behavioral constraints.

\begin{algorithm}[ht!]
\caption{Cross Evaluation \& Fitness Calculation}
\label{alg:cross-eval}
\begin{algorithmic}[1]
\Function{CrossEvaluation}{$\mathcal{P}_{\text{code}}, \mathcal{P}_{\text{test}}, R$}
    \State $M \gets \text{ZeroMatrix}(|\mathcal{P}_{\text{code}}|, |\mathcal{P}_{\text{test}}|)$
    \State Initialize maps $F_{\text{code}}, F_{\text{test}}$

    \State \textbf{Step 1: Execution Matrix Construction}
    \For{$c \in \mathcal{P}_{\text{code}}, t \in \mathcal{P}_{\text{test}}$}
        \State $res \gets \texttt{Run}(R \oplus c, t)$
        \State $M[c, t] \gets \mathbb{I}(res = \text{PASS})$
    \EndFor

    \State \textbf{Step 2: Compute Fitness}
    \For{$c \in \mathcal{P}_{\text{code}}$}
        \State $S_{\text{pass}} \gets \{t \mid M[c, t] = 1\}$ 
        \State $S_{\text{consensus}} \gets \{c' \in \mathcal{P}_{\text{code}} \mid M[c', \cdot] = M[c, \cdot]\}$
        \State $F_{\text{code}}[c] \gets |S_{\text{consensus}}| \times |S_{\text{pass}}|$
    \EndFor

    \For{$t \in \mathcal{P}_{\text{test}}$}
        \State $F_{\text{test}}[t] \gets \sum_{c \in \mathcal{P}_{\text{code}}} M[c, t] \cdot F_{\text{code}}[c]$
    \EndFor

    \State \Return $(F_{\text{code}}, F_{\text{test}})$
\EndFunction
\end{algorithmic}
\end{algorithm}

\noindent \textbf{Execution Matrix Construction:}
We initialize a binary matrix $M$ with dimensions $|\mathcal{P}_{\text{code}}| \times |\mathcal{P}_{\text{test}}|$.
For every code patch $c$ and test patch $t$, the system applies the patch to the repository, denoted as $R \oplus c$, and executes the test $t$.
The entry $M[c, t]$ is assigned a value of $1$ if the execution result is \texttt{PASS}, and $0$ otherwise.

\noindent \textbf{Fitness Calculation:} We employ a consensus-based metric to guide the evolution of both populations. 
First, for each code patch $c$,  we identify the set of passing tests $S_{\text{pass}}$. Simultaneously, we group code patches that exhibit identical behavior vectors into 
a set $S_{\text{consensus}}$. The fitness of a code patch is defined as:
\begin{equation}
F_{\text{code}}[c] = |S_{\text{consensus}}| \times |S_{\text{pass}}|.
\end{equation}
This formulation rewards patches that satisfy many informative tests while aligning with a stable behavioral consensus across the population.
Second, the fitness of a test patch $t$ is derived from the quality of the code patches it validates: 
\begin{equation}
F_{\text{test}}[t] = \sum_{c \in \mathcal{P}_{\text{code}}} M[c, t] \cdot F_{\text{code}}[c].
\end{equation}
This design prioritizes tests that are passed by higher-quality code hypotheses, thereby encouraging the evolution of constraints that better discriminate between promising and weak repair candidates.

\subsubsection{LLM-based Semantic Crossover}

To generate the next generation, we employ a semantic crossover operator.
Instead of syntactic splicing, the LLM analyzes the logical content of two parent patches and synthesizes an offspring that integrates their complementary strengths.
This operator enables aggregation of partial behavioral corrections discovered across different search trajectories.

\noindent \textbf{Parent Selection:} We utilize the Binary Tournament Selection method.
For each crossover operation, two candidates are randomly sampled from the population.
The individual with the higher fitness value is selected as a parent.
This process is repeated to select a second parent.
This strategy maintains a balance between population diversity and selective pressure.

\noindent \textbf{Semantic Synthesis:}
As shown in Phase 5 of Algorithm \ref{alg:coevolution-main}, the agents generate offspring sets.
The crossover function accepts two parents, $p_1$ and $p_2$.
Instead of splicing code strings, we construct a prompt that includes the semantic content of both patches.
The LLM analyzes the logic of $p_1$ and $p_2$ to synthesize a new patch that combines their strengths.

\subsubsection{Elite Reserve Strategy}
To guarantee the stability of the evolutionary process, we implement an elite reservation mechanism (Algorithm \ref{alg:coevolution-main}, Phase 4).
Before generating the new population, we identify the single best individual from the current generation, denoted as $e_{\text{code}}$ and $e_{\text{test}}$.
These elite individuals are directly transferred to the next generation, ensuring that the maximum fitness of the population is monotonically non-decreasing over iterations.
The remaining $N-1$ slots in the new population are filled by the offspring generated through crossover.

\subsubsection{Isolated Docker Environment Support}
To support the execution function $\texttt{Run}(R, t)$, we deploy a set of isolated Docker-based tools.
The Bash Tool provides an interface for executing shell commands, allowing the agents to run standard test suites and navigate the file system. 
The File Tool enables precise manipulation of the repository $R$, supporting operations such as creating files, patching content, and reverting changes to restore the initial state after each evaluation.
The ASTSearch and RegexSearch tools assist in the initial analysis by retrieving code contexts based on syntactic structures or text patterns.
This isolation ensures that the side effects of one patch evaluation do not contaminate subsequent evaluations.

In summary, \methodname casts repository-level repair into a coevolutionary search over implementation and behavioral constraints.
By coupling CodeAgent and TestAgent through mutual evaluation and refinement, the framework progressively reconciles code modifications with evolving test-based characterizations of intended behavior.

\section{Experimental Setup}
\subsection{Research Questions}
To evaluate the effectiveness of our coevolutionary framework of repository-level repair, we investigate the following research questions (RQs):
\begin{itemize}
\item \textbf{RQ1:} Does formulating repository-level issue resolution as the joint coevolution of code and tests yield measurable improvements over both code-only and test-only paradigms under comparable backbone models?
\item \textbf{RQ2:} How does the iterative coevolution process affect the effectiveness of generated code patches and test patches over successive iterations?
\item \textbf{RQ3:} What roles do key components, such as the coevolutionary mechanism, semantic crossover, and elite reservation, play in enabling the alignment between code hypotheses and evolving behavioral constraints?
\end{itemize}

\subsection{Benchmarks and Metrics}

We conduct our evaluation on the SWE-bench Lite \cite{jimenez2023swe} and SWT-bench Lite \cite{mundler2024swt}  datasets, which together provide complementary perspectives on repository-level issue resolution.
SWE-bench Lite is a lightweight subset of SWE-bench, containing 300 real-world software engineering issues from popular Python repositories on GitHub. 
Each instance provides a natural language issue description and the full repository snapshot at the buggy version. The task is to locate the faulty code and synthesize a patch that restores correct behavior. 
Evaluation is performed using a hidden test suite that includes both existing regression tests and issue-specific reproduction tests. 
A patch is considered successful if it passes all associated tests, and performance is measured using the resolution rate, defined as the percentage of issues correctly resolved.
While SWE-bench Lite adopts a standard test-based evaluation protocol, these hidden tests function as evaluation-time oracles rather than behavioral knowledge available during repair. 
In real-world issue resolution, such constraints must be iteratively constructed and refined, which aligns with our formulation that treats tests as evolving operational constraints approximating intended behavior.

SWT-bench Lite, derived from SWE-bench Lite, focuses on evaluating the ability to construct effective issue-reproducing tests. It contains 276 instances and uses the developer-provided golden patch as a reference to validate generated tests. A synthesized test is considered correct if it (1) fails on the original buggy repository and (2) passes after applying the golden fix. This setting evaluates how well generated tests capture the behavioral discrepancy underlying the issue.
From the perspective of our framework, SWT-bench Lite provides an indirect measure of the quality of evolved behavioral constraints. In addition to resolution rate, SWT-bench Lite reports the mean line coverage of the modified regions, 
denoted as  $\Delta \mathcal{C}$, which reflects how well the generated tests exercise the relevant fix locations.
In our experiments, we report resolution rates on both datasets and the $\Delta \mathcal{C}$ metric on SWT-bench Lite.

\subsection{Baselines}

We compare our framework against a broad set of state-of-the-art systems under comparable model capabilities. These baselines represent different methodological paradigms in automated repository-level software maintenance, including code-centric repair agents, test-generation agents, and generalist software agents.

The first group consists of \textbf{code-centric repair agents}, which focus on synthesizing code patches under fixed test validation. This category includes Agentless \cite{xia2024agentless}, SpecRover \cite{ruan2024specrover}, Moatless Tools \cite{orwall2024moatless}, SWE-Search \cite{antoniades2024swe}, KGCompass \cite{ma2025thinking}, and DARS \cite{aggarwal2025dars}. These methods primarily optimize code patches while treating tests as static acceptance criteria.
The second group comprises \textbf{test-generation agents}, which prioritize constructing reproduction or validation tests that capture failure behaviors. Representative methods include SWE-Agent+ \cite{mundler2024swt} and AssertFlip \cite{khatib2025assertflip}. These systems focus on modeling behavioral constraints but do not jointly optimize code modifications.
The third group includes \textbf{generalist software agents} capable of producing both code and tests within unified workflows, such as AutoCodeRover \cite{zhang2024autocoderover}, SWE-Agent \cite{yang2024swe}, and OpenHands \cite{wang2024openhands}. While these systems can generate multiple artifact types, they do not explicitly model repair as a coupled search over evolving behavioral constraints.

Our proposed method, \methodname, differs from all three paradigms by explicitly formulating repository-level repair as a coevolutionary process in which code patches and test patches are treated as interdependent search variables rather than artifacts optimized in isolation.
Due to the high computational cost and complex environment dependencies of repository-level repair systems, re-running all baselines under identical configurations is often impractical. Following prior SWE-bench studies \cite{ma2025thinking,aggarwal2025dars}, we report official leaderboard results or numbers from the original papers. These results reflect standard evaluation settings used in the literature. Our comparisons focus on methods operating within similar tiers of model capability rather than strict prompt-level replication.

For \methodname, the population size is set to 10, and the maximum number of evolutionary iterations is 5. The LocationAgent operates with a low-temperature setting (0.2) to promote stable localization, while the CodeAgent and TestAgent use a temperature of 0.5 to maintain generative diversity. All experiments are conducted using DeepSeek-V3-0324 \cite{liu2024deepseek} via OpenRouter \footnote{https://openrouter.ai/} API access.

\section{Results}

\subsection{RQ1: Performance Comparison}
We evaluate \methodname against competitive baselines powered by comparable LLMs on SWE-bench Lite and SWT-bench Lite. Our goal is to assess whether modeling repository-level repair as a coevolution process leads to measurable gains under similar foundation model capabilities.

\begin{table}[h!]
\caption{
Performance comparison of \methodname and baseline methods on SWE-bench Lite and SWT-bench Lite. 
Resolved indicates the percentage of issues successfully fixed. $\Delta \mathcal{C}$ represents the improvement in test consistency and coverage. Best results are highlighted in bold.
}
\label{tab:rq1}
\resizebox{\textwidth}{!}{

\begin{tabular}{llcccc}
\toprule
\multirow{2}{*}{\textbf{Method}} & \multirow{2}{*}{\textbf{LLM}} & \multicolumn{1}{c}{\textbf{SWE-bench Lite (300)}} & & \multicolumn{2}{c}{\textbf{SWT-bench Lite (276)}} \\

\cmidrule{3-3}
\cmidrule{5-6}

&& \textbf{Resolved (\%)} && \textbf{Resolved (\%)} & \textbf{$\Delta \mathcal{C}$} \\

\midrule

Agentless \cite{xia2024agentless} 
    & GPT-4o            & 27.33\% && $-$ & $-$ \\

Agentless 1.5 \cite{xia2024agentless}
    & GPT-4o            & 32.00\% && $-$ & $-$ \\

OpenHands v1.8 \cite{wang2024openhands}
    & GPT-4o-mini           & 6.33\%   && $-$       & $-$ \\
    & GPT-4o                & 22.00\%  && $-$       & $-$ \\
    & Claude 3.5 Sonnet     & 26.00\%  && $-$       & $-$ \\

SpecRover \cite{ruan2024specrover} 
    & Claude 3.5 Sonnet+GPT-4o & 31.00\% && $-$ & $-$ \\

Moatless Tools \cite{orwall2024moatless} 
    & DeepSeek-V3           & 30.67\% && $-$ & $-$ \\
    & Claude 3.5 Sonnet     & 39.00\% && $-$ & $-$ \\

SWE-Search \cite{antoniades2024swe} 
    & GPT-4o        & 31.00\% && $-$ & $-$ \\

KGCompass \cite{ma2025thinking} 
    & DeepSeek-V3   & 36.67\%     && $-$ & $-$ \\

DARS \cite{aggarwal2025dars}
    & GPT-4o        & 37.00\%     && $-$ & $-$ \\

\midrule

LIBRO \cite{kang2023large}
    & GPT-4     & $-$ && 14.1\% & 23.8\% \\

SWE-Agent+ \cite{mundler2024swt}
    & GPT-4     & $-$ && 18.5\% & 27.6\% \\

OpenHands vanilla \cite{mundler2024swt}
    & Claude 3.5 Sonnet     & $-$       && 22.8\%  & 43.6\% \\

OpenHands setup \cite{mundler2024swt}
    & Claude 3.5 Sonnet     & $-$       && 28.3\%  & 52.4\% \\

AssertFlip \cite{khatib2025assertflip} 
    & GPT-4o    & $-$ && 38.0\% & 44.2\% \\

AEGIS \cite{wang2024aegis} 
    & GPT-4o    & $-$ && 36.0\% & 44.2\% \\

\midrule

AutoCodeRover \cite{zhang2024autocoderover}
    & GPT-4     & 19.00\%  && 9.1\%     & 17.9\%  \\
    & GPT-4o    & 30.67\%  && $-$       & $-$     \\

SWE-Agent \cite{yang2024swe}
    & GPT-4                 & 18.00\%   && 15.9\%   & 26.5\% \\
    & GPT-4o mini           & $-$       && 9.8\%    & 20.9\% \\
    & GPT-4o                & 18.33\%   && $-$      & $-$ \\
    & Claude 3.5 Sonnet     & $-$       && 12.3\%   & 30.3\% \\

\textbf{\methodname} 
    & \textbf{DeepSeek-V3}  & \textbf{41.33\%} && \textbf{46.4\%} & \textbf{56.0\%} \\

\bottomrule
\end{tabular}
}
\end{table}

The results in Table~\ref{tab:rq1} show that \methodname achieves the best performance on both benchmarks. 
On SWE-bench Lite, \methodname attains a resolution rate of 41.33\%, exceeding strong code-centric repair agents such as DARS (37.00\%) and KGCompass (36.67\%). 
On SWT-bench Lite, which evaluates the ability to construct behaviorally correct reproduction tests, \methodname reaches 46.4\%, substantially outperforming prior test-generation systems such as AssertFlip (38.0\%). 
Additionally, \methodname achieves the highest $\Delta \mathcal{C}$ score (56.0\%), indicating that the synthesized tests better capture the behavior of the reference fixes.

These results are notable because the baselines represent fundamentally different solution paradigms: code-centric repair agents optimize code under fixed test validation, while test-generation agents focus on modeling failure behavior without jointly optimizing code. In contrast, \methodname treats both code patches and tests as interdependent search variables. The consistent improvements across both SWE-bench Lite (repair correctness) and SWT-bench Lite (behavioral constraint quality) suggest that joint optimization over implementation and constraints is more effective than optimizing either component in isolation.

\begin{figure}[h!]
\centering
\begin{minipage}{0.55\textwidth}
\centering
\includegraphics[width=\textwidth]{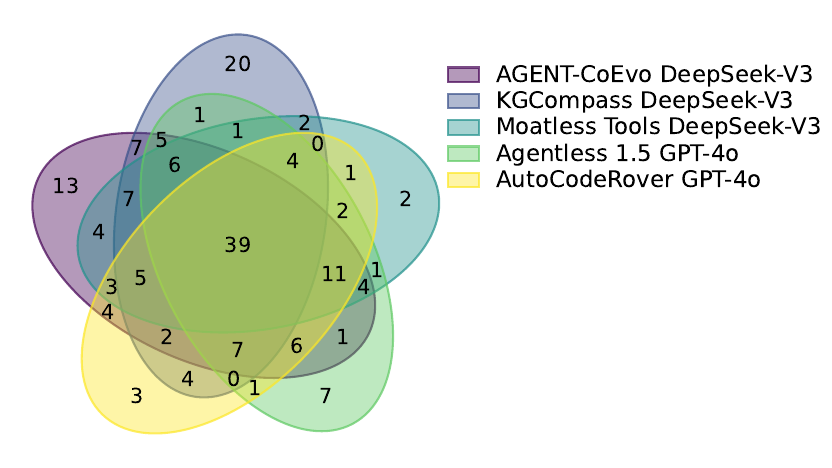}
\caption{
Venn diagram illustrating the overlap of resolved issues. 
Numbers indicate the specific count of issues in each intersection.
}
\label{fig:venn}
\end{minipage}
\hfill
\begin{minipage}{0.42\textwidth}
\centering
\captionof{table}{
Comparison of issue coverage. 
Unique counts issues solved exclusively by each method, while UR (Union Recall) quantifies the coverage of the collective solution space.
}
\label{tab:coverage}
\small
\resizebox{\textwidth}{!}{

\begin{tabular}{lccc}
\toprule
\textbf{Method} & \textbf{Count} & \textbf{Unique} & \textbf{UR (\%)} \\
\midrule
\textbf{\methodname} & \textbf{124} & 13 & \textbf{71.68} \\
KGCompass & 110 & \textbf{20} & 63.58 \\
Agentless 1.5 & 96 & 7 & 55.49 \\
Moatless Tools & 92 & 2 & 53.18 \\
AutoCodeRover & 92 & 3 & 53.18 \\
\bottomrule
\end{tabular}
}
\end{minipage}
\end{figure}

\paragraph{Analysis of Overlap and Robustness.}
To better understand the nature of these gains, we analyze the overlap between issues solved by \methodname and leading baselines (KGCompass, Moatless Tools, Agentless, and AutoCodeRover), shown in Figure~\ref{fig:venn}.
Complementing the visual representation, Table~\ref{tab:coverage} details the quantitative performance regarding specific issue counts and the Union Recall (UR). 
We define $\mathcal{S}_m$ as the set of issues successfully resolved by a specific method $m$. Let $\mathcal{U}$ denote the union of all distinct issues resolved by the combination of \methodname and the top baseline methods, such that $\mathcal{U} = \bigcup_{k} \mathcal{S}_k$. The UR for method $m$ quantifies the coverage of the collective solution space and is calculated by:
\begin{equation}
\label{eq:ur}
UR_m = \frac{|\mathcal{S}_m|}{|\mathcal{U}|} \times 100\%
\end{equation}
This metric demonstrates the extent to which a single approach captures the total problem set solvable by current state-of-the-art capabilities.

The overlap patterns reveal three key properties:
\begin{enumerate}
\item \textbf{Core Robustness:} A substantial subset of issues (39 cases) is solved by all top-performing methods. \methodname retains full coverage of this consensus set, indicating that the coevolution formulation does not sacrifice performance on problems already accessible to existing paradigms.
\item \textbf{Cross-Paradigm Coverage:} \methodname covers most issues solved by methods built on distinct mechanisms (e.g., knowledge-graph retrieval in KGCompass and structured tool pipelines in Agentless and Moatless). This suggests that the coevolution search process can subsume solution patterns discovered by multiple paradigms.
\item \textbf{Unique Resolution Capability:} In terms of exclusive coverage, \methodname ranks second only to KGCompass, resolving 13 unique issues. This count is nearly double that of the next best method (Agentless). These cases specifically require both correcting implementation logic and refining behavioral constraints, a capability lacking in methods that treat tests as fixed filters or generate them independently.
\end{enumerate}

Overall, the performance gains of \methodname arise not from solving isolated niche cases, but from reducing false negatives caused by mismatches between code hypotheses and imperfect behavioral constraints. By allowing tests to evolve together with code patches, the system can recover from early mis-specifications and converge toward behaviorally consistent solutions.

\begin{AnswerBox}
\textbf{Answer to RQ1:}
\methodname achieves state-of-the-art performance on both SWE-bench Lite (\textbf{41.33\%}) and SWT-bench Lite (\textbf{46.4\%}), surpassing existing code-centric and test-centric baselines under comparable backbone models. 
Overlap analysis shows that \methodname preserves all consensus-solvable cases while uniquely resolving \textbf{13} additional issues, demonstrating that co-evolving code and tests reduces false negatives introduced by fixed or imperfect constraints and enables recovery from early search errors.
\end{AnswerBox}

\subsection{RQ2: Impact of Evolutionary Iterations}

In this section, we investigate how iterative coevolution influences the progressive alignment between generated code patches and behavioral constraints. 
Unlike conventional search procedures where additional iterations mainly increase sampling opportunities, 
our framework is designed such that successive generations refine both implementation hypotheses and constraint quality through cross-population feedback.

\begin{table}[h!]
\caption{
Impact of the number of evolution iterations on performance and cost. The cost represents the average dollar expenditure per issue.
}
\label{tab:rq2}
\begin{tabular}{cccc}
\toprule
\textbf{Iteration} & \textbf{SWE-bench Lite (300)} & \textbf{SWT-bench Lite (276)} & \textbf{Avg. Cost (\$)} \\
\midrule
1           & 33.67\%           & 42.4\%            & 0.32          \\
2           & 36.00\%           & 42.4\%            & 0.54          \\
3           & 39.33\%           & 44.9\%            & 0.74          \\
4           & 39.67\%           & 44.2\%            & 0.93          \\
\textbf{5}  & \textbf{41.33\%}  & \textbf{46.4\%}   & \textbf{1.11} \\
\bottomrule
\end{tabular}
\end{table}

\begin{figure}[h!]
\centering
\includegraphics[width=0.6\linewidth]{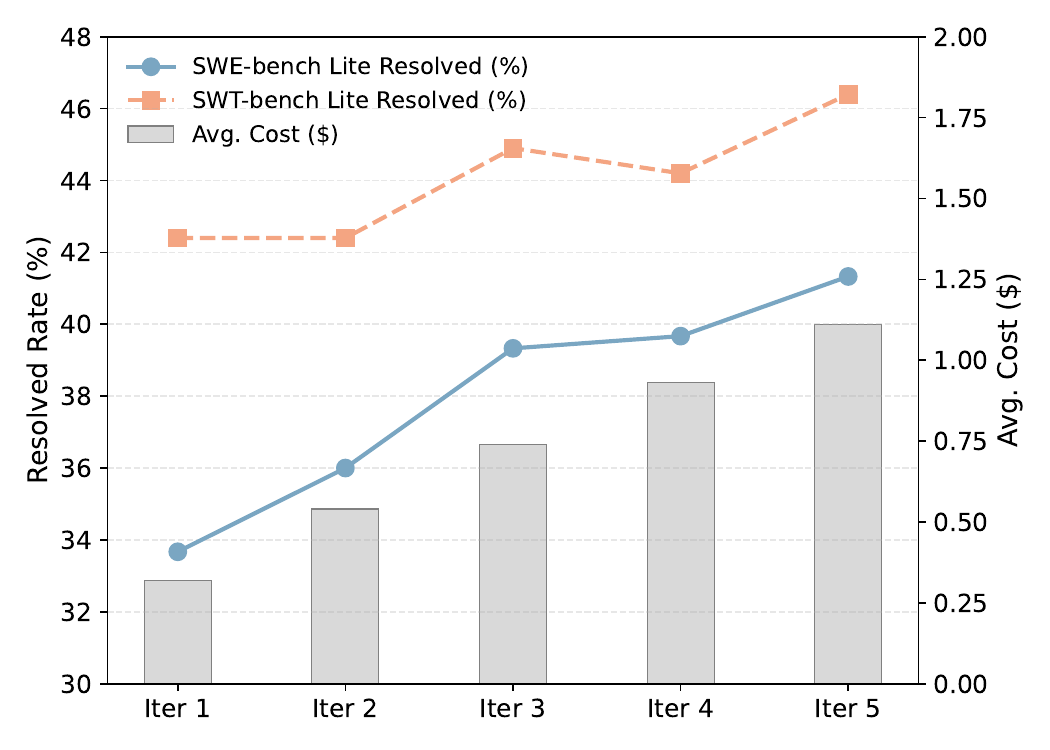}
\caption{
Evaluation of resolved rates and average costs across five evolutionary iterations on SWE-bench Lite and SWT-bench Lite.
}
\label{fig:iters}
\Description{
The chart plots the resolved rate and average cost over 5 iterations. The resolved rates show an upward trend, peaking at iteration 5. The cost increases linearly.
}
\end{figure}

As shown in Table~\ref{tab:rq2} and Figure~\ref{fig:iters}, repair performance improves steadily across iterations on both benchmarks. On SWE-bench Lite, the resolution rate increases from 33.67\% at Iteration 1 to 41.33\% at Iteration 5. A similar upward trend is observed on SWT-bench Lite. These results indicate that the coevolutionary process does not merely expand the search space, but progressively enhances behavioral consistency between candidate code patches and evolving test constraints.

A slight fluctuation appears at Iteration 4 on SWT-bench Lite. Such variations are expected in population-based coevolution, where newly introduced variants may temporarily alter the behavioral landscape. Importantly, performance recovers in Iteration 5, suggesting that the search process exhibits self-correcting dynamics rather than being trapped in unstable states. This behavior is consistent with our design, where cross-population evaluation gradually filters inconsistent hypotheses while reinforcing mutually compatible code–test pairs.

From a practical perspective, computational cost grows approximately linearly with the number of iterations. Iteration 3 offers a favorable trade-off between effectiveness and cost, while Iteration 5 provides the strongest repair capability. Beyond this point, additional iterations would likely yield diminishing returns relative to their cost.

\begin{AnswerBox}
\textbf{Answer to RQ2:}
Iterative coevolution leads to progressive behavioral alignment between code patches and evolving test constraints. Resolution rates improve across iterations, indicating that later iterations refine constraint quality rather than merely increasing search attempts. Although minor fluctuations arise from population dynamics, the process exhibits self-correcting behavior and reaches peak performance in later iterations.
\end{AnswerBox}

\subsection{RQ3: Ablation Study}

To understand how different components contribute to the coevolutionary alignment process, we conduct an ablation study on \methodname. Specifically, we remove key mechanisms that govern the interaction between code hypotheses and behavioral constraints, and evaluate how each removal affects repair performance.
The following variants are considered:
\begin{itemize}
\item ``w/o TestAgent'': removes the evolving test population. Code patches are still generated and evaluated, but behavioral constraints no longer evolve and reduce to static validation signals.
\item ``w/o Evolution'': removes semantic crossover and offspring generation. The system retains initialization and cross-evaluation but degenerates into independent local search without hypothesis recombination.
\item ``w/o Elite Reserve'': removes the memory-preserving mechanism. Parent individuals are discarded after each generation, preventing stable retention of previously aligned code–test pairs.
\end{itemize}
To control computational cost, we fix the number of evolutionary iterations to 3 for all variants. Experiments are conducted using DeepSeek-V3 on SWE-bench Lite and SWT-bench Lite. Results are shown in Table~\ref{tab:rq3}.

\begin{table}[h!]
\caption{
Ablation study of \methodname on the SWE-bench Lite and SWT-bench Lite datasets using DeepSeek-V3.
}
\label{tab:rq3}
\begin{tabular}{lcccc}
\toprule

\textbf{Method (Iter 3)} & \textbf{SWE-bench Lite (300)} & \textbf{SWT-bench Lite (276)} \\
\midrule
\textbf{\methodname}  & \textbf{39.33\%}  & \textbf{44.9\%}   \\
w/o TestAgent           & 33.33\%           & $-$      \\
w/o Evolution           & 33.67\%           & 42.4\%   \\
w/o Elite Reserve       & 37.33\%           & 39.1\%   \\
\bottomrule
\end{tabular}
\end{table}

The ablation results show consistent performance degradation across all variants, indicating that each component contributes to overall repair effectiveness. Removing the TestAgent leads to the largest drop on SWE-bench Lite (39.33\% $\rightarrow$ 33.33\%), suggesting that the ability to refine behavioral feedback during search is important for successful repair. Disabling semantic evolution also reduces performance (39.33\% $\rightarrow$ 33.67\%), showing that recombination of partially correct hypotheses improves solution quality beyond independent sampling. Removing elite reservation causes a smaller but noticeable decline, especially on SWT-bench Lite (44.9\% $\rightarrow$ 39.1\%), indicating that retaining high-performing individuals helps maintain stability across generations.

Overall, the results suggest that the performance gains of \methodname arise from the coordinated interaction of evolving tests, hypothesis recombination, and memory retention, rather than from any single mechanism in isolation.

\begin{AnswerBox}
\textbf{Answer to RQ3:}
Performance gains arise from the coordinated alignment mechanism rather than any single module. Removing the TestAgent, semantic evolution, or elite reservation each degrades repair success, indicating that effective repair depends on the joint interaction between evolving behavioral feedback, hypothesis recombination, and cross-generation retention.
\end{AnswerBox}

\section{Discussion}

\subsection{Case Study: Efficacy of Semantic-Based Crossover}

\begin{figure}[h!]
\centering
\includegraphics[width=0.7\linewidth]{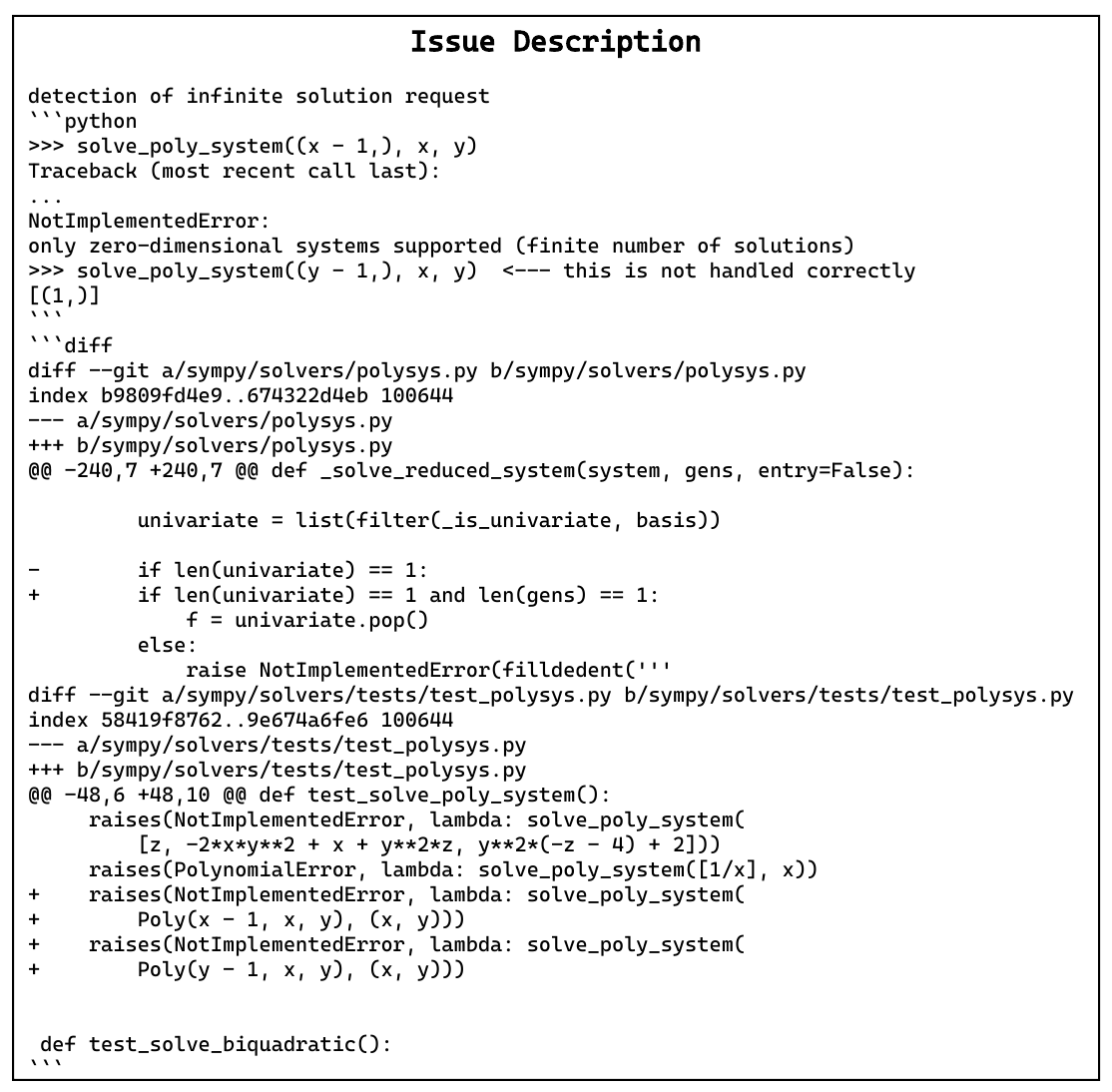}
\caption{The issue description.}
\label{fig:issue-description}
\Description{
}
\end{figure}

\begin{figure}[h!]
\centering
\includegraphics[width=\linewidth]{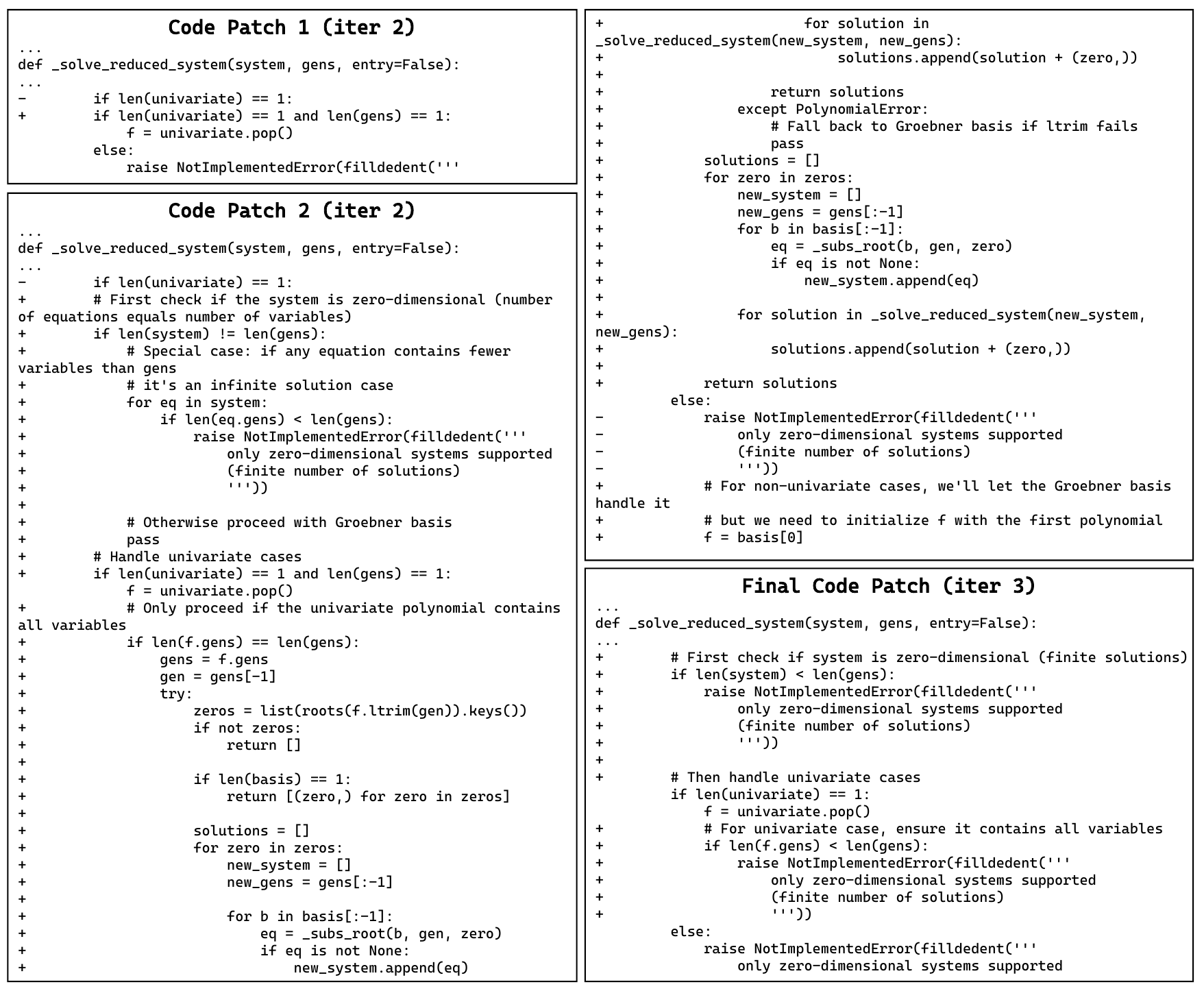}
\caption{
Evolution of code patches. Iteration 2 produces two partial hypotheses; Iteration 3 synthesizes them into an aligned solution via semantic crossover.
}
\label{fig:code-crossover}
\Description{

}
\end{figure}

\begin{figure}[h!]
\centering
\includegraphics[width=\linewidth]{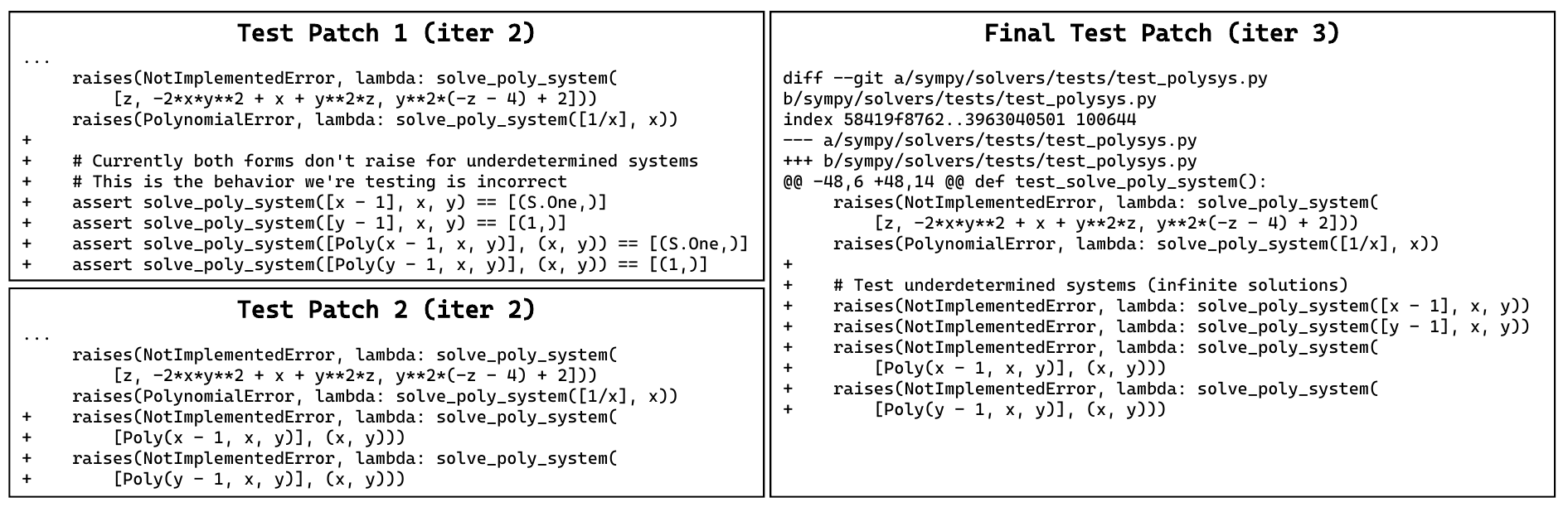}
\caption{
The refinement of test patches. The crossover mechanism aligns the test cases with the correct exception handling logic.
}
\label{fig:test-crossover}
\Description{}
\end{figure}

To understand why coevolution improves repair quality beyond simple candidate generation, we analyze a representative case from the SymPy repository (issue \verb|sympy__sympy-22005|). Figure~\ref{fig:issue-description} presents the original issue description, Figure~\ref{fig:code-crossover} illustrates the evolution and crossover of code patches, and Figure~\ref{fig:test-crossover} shows the corresponding refinement in the test population.

The issue concerns the \verb|solve_poly_system| function, which computes solutions for systems of polynomial equations. The defect arises from the absence of a dimensionality validation mechanism. When the number of equations is fewer than the number of variables, the system is underdetermined and admits infinitely many solutions. The original implementation failed to detect this condition and proceeded with recursive solving, leading to infinite recursion or incorrect results. The correct behavior is to enforce a zero-dimensionality constraint and raise a \texttt{NotImplementedError} when a finite solution set cannot be guaranteed.

During Iteration 2, the CodeAgent produced two partial but complementary hypotheses (Figure~\ref{fig:code-crossover}). \textbf{Patch 1} introduced a local guard in the univariate handling branch. This mitigated specific recursion failures but lacked a global understanding of system dimensionality, resulting in only partial behavioral correction. \textbf{Patch 2}, in contrast, correctly identified the global dimensionality condition by comparing the number of equations with the number of variables. However, its implementation was overly complex, introducing redundant Groebner basis computations and manual substitution logic, producing correct but bloated and hard-to-maintain code.
Neither patch alone represented a well-aligned solution: Patch 1 captured local behavioral stabilization, while Patch 2 captured global semantic correctness.

In Iteration 3, semantic crossover synthesized these hypotheses into a concise and correct patch. The resulting solution preserved the essential global dimensionality check \verb|len(system) < len(gens)| from Patch 2 while retaining the simpler recursion structure and contextual consistency checks inspired by Patch 1. Importantly, this synthesis was not a surface-level recombination but resulted from LLM-driven reasoning over the semantic roles of each patch. The final solution is both mathematically rigorous and structurally coherent.

A parallel alignment occurred in the test population (Figure~\ref{fig:test-crossover}). Early test patches encoded inconsistent behavioral interpretations, including incorrect expected outputs for underdetermined systems. Through coevolution, crossover aligned the test population toward consistent enforcement of the correct exception behavior (\texttt{NotImplementedError}). Thus, both code and test populations converged toward a shared behavioral interpretation of the issue.

This case demonstrates that semantic crossover functions as a hypothesis alignment mechanism. Rather than merely selecting the best individual, the process reconciles complementary behavioral interpretations across candidate patches, enabling convergence toward solutions that are simultaneously semantically correct and structurally coherent. The improvement arises from progressive alignment between implementation hypotheses and evolving behavioral constraints, 
which illustrates the core premise of this work: coevolution operates as an alignment process rather than merely a larger search.

\subsection{Cost Analysis: Structural Trade-off of Coupled Search}
Table \ref{tab:dis2} reports the average cost and performance of \methodname in comparison with representative baselines. While \methodname incurs a higher per-instance cost than some code-centric pipelines (e.g., Agentless, AutoCodeRover), its computational profile reflects a fundamentally different search structure rather than from redundant computation.

\begin{table}[h!]
\caption{Comparison of performance and cost across different methods.}
\label{tab:dis2}
\begin{tabular}{lcccc}
\toprule
\textbf{Method} & \textbf{LLM} & \textbf{SWE-bench Lite} & \textbf{SWT-bench Lite} & \textbf{Avg. Cost (\$)} \\
\midrule
Agentless
    & GPT-4o        & 27.33\%   & $-$       & 0.70  \\
AutoCodeRover
    & GPT-4         & 19.00\%   & $-$       & 0.47  \\
DARS
    & GPT-4o        & 37.00\%   & $-$       & 7.92  \\
SWE-Agent+
    & GPT-4         & $-$       & 18.5\%    & 1.73  \\
\methodname
    & DeepSeek-V3   & 41.33\%   & 46.4\%    & 1.11 \\

\bottomrule
\end{tabular}
\end{table}

Conventional repository-level repair methods operate primarily in the implementation space, treating tests as fixed validation filters. 
In contrast, \methodname performs coupled search over both code hypotheses and behavioral constraints by evolving test patches alongside code patches. 
The additional cost therefore stems from modeling a more complete repair loop that includes specification refinement, rather than from redundant inference.

This structural distinction explains the trends in Table \ref{tab:dis2}. Code-centric systems such as Agentless (27.33\%, \$0.70) and AutoCodeRover (19.00\%, \$0.47) achieve lower cost by restricting search to implementation hypotheses under static constraints, but this limits recovery when behavioral signals are incomplete. In comparison, \methodname (41.33\%, \$1.11) achieves substantially higher repair success on SWE-bench Lite and strong performance on SWT-bench Lite (46.4\%), where behavioral correctness is explicitly evaluated. The gain reflects the ability to refine constraints during search rather than relying on fixed tests.

From a practical perspective, \methodname simultaneously produces two artifacts: a code patch and its behavioral verification (in the form of test patch). In real maintenance workflows, implementation changes and validation co-evolve; our framework internalizes this interaction within the search process, enabling alignment between implementation hypotheses and evolving behavioral constraints.

Overall, the cost of \methodname represents a structural trade-off between search completeness and computational efficiency. By extending search from the implementation space to a coupled implementation–constraint space, the framework incurs additional computation but achieves greater reliability while jointly producing a fix and its behavioral verification. This structural trade-off also points to future research on adaptive strategies that balance search completeness and computational cost.

\section{Limitations}

Despite its strong performance, the proposed framework has several limitations.
First, the coevolutionary formulation incurs higher computational cost than code-centric pipelines because it performs coupled search over both implementation hypotheses and behavioral constraints. While this improves repair reliability, it may introduce unnecessary overhead for simple defects; future work could explore adaptive strategies that switch between lightweight repair and full coevolution based on issue complexity.
Second, the behavioral constraints themselves are learned signals rather than perfect specifications. LLM-generated tests may contain subtle semantic inaccuracies \cite{ahmed2024tdd,ahmed2025otter,mundler2024swt}, which can influence the optimization landscape despite our filtering mechanisms.
Finally, the effectiveness of coevolution depends on initial fault localization; inaccurate localization may restrict the hypothesis space and affect early search trajectories. Enhancing structural code understanding for localization \cite{ma2025thinking}
remains an important direction for improving alignment-driven repair systems.

\section{Threats to Validity}

\paragraph{Internal Validity.} Internal threats mainly stem from stochasticity in LLM generation and evolutionary sampling, which may influence population diversity and search trajectories. 
We mitigate this by fixing iteration settings, using consistent evaluation protocols, and applying identical backbone models across comparisons, ensuring that performance differences arise from methodological design rather than random variation.

\paragraph{External Validity.} Our evaluation focuses on Python-based repositories, which may limit generalization to other programming ecosystems. 
Although the coevolution framework is conceptually language-agnostic, practical deployment depends on language-specific build systems, execution environments, and test infrastructures.

\paragraph{Construct Validity.} Construct validity is affected by the reliance on test-based evaluation. Passing test suites reflects behavioral consistency under available constraints rather than full semantic equivalence to developer fixes. In our framework, tests are treated as evolving behavioral constraints, while the benchmark-provided evaluation test suites serve as the external oracle. However, these benchmark tests still approximate intended behavior rather than constituting complete formal specifications.

\section{Related Work}

\subsection{Repository-Level Issue Resolution}

Repository-level issue resolution has become a central testbed for evaluating LLMs in realistic software engineering settings. The introduction of SWE-bench \cite{jimenez2023swe} established an executable benchmark grounded in real GitHub issues with hidden test-based evaluation, revealing a substantial gap between surface-level code generation and the robustness required for practical maintenance. Subsequent analyses further show that many reported “solutions” remain incorrect or incomplete \cite{wang2025solved}, emphasizing the need for stronger repository reasoning and reliable evaluation.

Most existing repository-level repair approaches operate under a \emph{fixed-constraint repair paradigm}, where validation signals remain static during repair, typically derived from existing tests, regression suites, or basic correctness checks. Under this setting, the search process primarily explores the space of code patches that satisfy these fixed behavioral signals.
A large body of work explores agent-based repository reasoning and patch synthesis under this assumption. Systems such as SWE-agent \cite{yang2024swe}, OpenHands \cite{wang2024openhands}, MarsCode \cite{liu2024marscode}, and AutoCodeRover \cite{zhang2024autocoderover} design tool-augmented agents to navigate codebases and iteratively refine patches. Other efforts investigate structured collaboration \cite{chen2024coder}, historical and dependency-aware exploration \cite{ma2025alibaba,ma2024lingma}, or simplified agent pipelines that reduce step-wise error propagation \cite{xia2024agentless}.

Another line of work improves repair effectiveness by enhancing repository understanding and structured reasoning. Methods such as RepoGraph \cite{ouyang2024repograph}, knowledge-graph–augmented approaches \cite{yang2025enhancing}, and intent-grounding systems like SpecRover \cite{ruan2024specrover} inject structural and semantic information into the repair process. Search-driven techniques including SWE-Search \cite{antoniades2024swe} and multi-agent debate frameworks \cite{li2025swe} further guide patch exploration, while experience-based systems \cite{chen2025swe,mu2025experepair} leverage memory to transfer prior repair knowledge across tasks.

Recent studies also highlight the importance of inference-time compute scaling. Approaches such as Thinking Longer, Not Larger \cite{ma2025thinking} and Trae Agent \cite{gao2025trae} show that deeper test-time reasoning and reflection substantially improve repair success. Complementary work like BugPilot \cite{sonwane2025bugpilot} focuses on constructing more complex bug scenarios to stress-test these systems.

Across these directions, the common assumption is that behavioral constraints are externally provided and remain fixed during repair, 
and progress primarily comes from improving repository navigation, reasoning depth, or search strategies within this constraint space.
In contrast, our work departs from this paradigm by treating behavioral constraints themselves as evolving entities during repair, 
enabling a coupled search over both implementation hypotheses and constraint formulations.

\subsection{Repository-Level Issue Reproduction}

Repository-level issue reproduction focuses on translating natural language issue reports into executable artifacts that trigger the observed faulty behavior. The introduction of SWT-bench \cite{mundler2024swt} established a benchmark for evaluating this capability, emphasizing the challenge of bridging high-level problem descriptions and concrete test scenarios.

Early work largely framed reproduction as a direct generation task. Methods such as LIBRO \cite{kang2023large}, Issue2Test \cite{nashid2025issue2test}, and AssertFlip \cite{khatib2025assertflip} prompt LLMs to synthesize reproduction scripts or failing tests from issue descriptions, sometimes through intermediate transformations (e.g., generating passing tests before assertion inversion). In these approaches, tests primarily serve as artifacts that demonstrate failure, and generation quality depends heavily on prompt design and language understanding.

To improve reliability, subsequent research incorporates execution feedback and search-based refinement. Systems like e-Otter+ \cite{ahmed2025heterogeneous} and hybrid LLM–SBST frameworks \cite{kitsios2025automated} iteratively adjust tests using runtime signals and coverage information. These methods treat reproduction as an optimization process over test inputs, but the produced tests remain validation endpoints rather than dynamic components of a broader repair loop.

More recently, agent-based frameworks such as AEGIS \cite{wang2024aegis} and general-purpose software agents (e.g., SWE-agent \cite{yang2024swe}, OpenHands \cite{wang2024openhands}, AutoCodeRover \cite{zhang2024autocoderover}) incorporate test reproduction into larger repository reasoning pipelines. Here, test generation becomes part of a multi-step workflow that includes navigation, reasoning, and environment interaction. Nevertheless, across these approaches, tests are typically generated to reproduce or validate issues, while remaining structurally separate from the repair search process.

Overall, existing issue reproduction research treats tests as generated validation artifacts rather than as variables that co-evolve with code during repair.
In contrast, our framework models tests as evolving behavioral constraints that co-evolve with code hypotheses, integrating validation directly into the repair dynamics.

\subsection{Feedback-Driven Iterative Repair}

Iterative refinement guided by execution feedback has become a central paradigm for improving the correctness and reliability of LLM-generated code.
Early work on model self-debugging~\cite{chen2023teaching} showed that LLMs can leverage compiler or runtime error messages to analyze failures and propose targeted revisions, establishing the importance of grounding refinement in actual program behavior. This idea was later generalized in Self-Refine~\cite{madaan2023self}, which introduced a task-agnostic feedback loop where model outputs are repeatedly critiqued and improved. Building on these concepts, Reflexion~\cite{shinn2023reflexion} incorporates verbal self-reflection to accumulate reasoning strategies across iterations, while RLEF~\cite{gehring2024rlef} uses reinforcement learning from execution signals to directly shape repair policies.

A complementary direction enriches feedback-driven repair by extending the loop beyond code-only refinement. 
LEVER~\cite{ni2023lever} emphasizes verification-in-the-loop through constraint checks and invariant validation, 
while subsequent work revisits self-debugging through self-generated tests~\cite{chen2025revisit}, showing that adaptive test synthesis can improve fault localization. 
RepairAgent~\cite{bouzenia2024repairagent} further operationalizes iterative feedback within autonomous agents that integrate code navigation, environment interaction, and repeated refinement for real repair tasks.
More recently, several systems explicitly model interaction between candidate programs and generated tests during refinement. Representative examples such as LLMLOOP~\cite{ravi2025llmloop} and CoCoEvo~\cite{li2025cocoevo} 
jointly generate tests and program candidates, enabling mutual influence between solutions and their evaluators within a unified generation framework.
However, these systems are primarily studied in function-level code synthesis settings, where target behavior is implicitly defined, whereas repository-level issue resolution
requires jointly repairing implementations and refining the behavioral constraints themselves.

Overall, prior feedback-driven repair research primarily treats execution signals as evaluative feedback used to improve candidate implementations. 
In contrast, our work views repository-level issue resolution as search over both implementations and the behavioral constraints, where tests are not only feedback signals 
but evolving representations of partially known specifications.

\section{Conclusion}

This work revisits the modeling assumption underlying repository-level issue resolution. 
Rather than treating repair as optimization under fixed validation signals, we argue that real-world resolution involves searching over evolving behavioral constraints where both implementations and tests are iteratively refined.
To operationalize this, we proposed \methodname, a coevolutionary multi-agent framework that jointly explores code and test patches. 
Through mutual evaluation and semantic recombination, the system aligns implementation hypotheses with evolving constraints, allowing recovery from imperfect initial signals.
Evaluation on SWE-bench Lite and SWT-bench Lite shows that \methodname consistently outperforms competitive baselines under comparable backbone models.
Broadly, our findings advocate a shift in perspective for automated issue resolution: from code-only optimization under static tests to the coevolution of implementation and specification.
Future work may explore adaptive evolution strategies to balance search completeness and computational cost.


\bibliographystyle{ACM-Reference-Format}
\bibliography{paper}

\appendix

\end{document}